\documentstyle[12pt,epsfig,color]{article}
\def\baselinestretch{1.3}
\advance\textheight by \topskip
\newcommand{\comment}[1]{}

\def\beq{\begin{equation}}
\def\eeq{\end{equation}}
\def\beqn{\begin{eqnarray}}
\def\eeqn{\end{eqnarray}}

\begin{document}
 \tolerance=100000
 \topmargin -0.1in
\headsep 30pt
\footskip 40pt
\oddsidemargin 12pt
\evensidemargin -16pt
\textheight 8.5in
\textwidth 6.5in
\parindent 20pt
 
\def\baselinestretch{1.5}
\newcommand{\newc}{\newcommand}
\def\preprint{{preprint}}
\def\Ord{\lower .7ex\hbox{$\;\stackrel{\textstyle <}{\sim}\;$}}
\def\OOrd{\lower .7ex\hbox{$\;\stackrel{\textstyle >}{\sim}\;$}}
\def\cO#1{{\cal{O}}\left(#1\right)}
\newc{\order}{{\cal O}}
\def\lag             {{\cal L}}
\def\Lag             {{\cal L}}
\def\lum             {{\cal L}}
\def\R               {{\cal R}}
\def\Rsq             {{\cal R}^{\sq}}
\def\Rst             {{\cal R}^{\st}}
\def\Rsb             {{\cal R}^{\sb}}
\def\M               {{\cal M}}
\def\Oas             {{\cal O}(\alpha_{s})}
\def\Vcal            {{\cal V}}
\def\Wcal            {{\cal W}}
\newc{\be}{\begin{equation}}
\newc{\ee}{\end{equation}}
\newc{\br}{\begin{eqnarray}}
\newc{\er}{\end{eqnarray}}
\newc{\ba}{\begin{array}}
\newc{\ea}{\end{array}}
\newc{\bi}{\begin{itemize}}
\newc{\ei}{\end{itemize}}
\newc{\bn}{\begin{enumerate}}
\newc{\en}{\end{enumerate}}
\newc{\bc}{\begin{center}}
\newc{\ec}{\end{center}}
\newc{\ul}{\underline}
\newc{\ol}{\overline}
\newc{\ra}{\rightarrow}
\newc{\lra}{\longrightarrow}
\newc{\wt}{\widetilde}
\newc{\til}{\tilde}
\def\kr              {^{\dagger}}
\newc{\wh}{\widehat}
\newc{\ti}{\times}
\newc{\Dir}{\kern -6.4pt\Big{/}}
\newc{\Dirin}{\kern -10.4pt\Big{/}\kern 4.4pt}
\newc{\DDir}{\kern -10.6pt\Big{/}}
\newc{\DGir}{\kern -6.0pt\Big{/}}
\newc{\sig}{\sigma}
\newc{\sigmalstop}{\sig_{\lstoppair}}
\newc{\Sig}{\Sigma}  
\newc{\del}{\delta}
\newc{\Del}{\Delta}
\newc{\lam}{\lambda}
\newc{\Lam}{\Lambda}
\newc{\gam}{\gamma}
\newc{\Gam}{\Gamma}
\newc{\eps}{\epsilon}
\newc{\Eps}{\Epsilon}
\newc{\kap}{\kappa}
\newc{\Kap}{\Kappa}
\newc{\modulus}[1]{\left| #1 \right|}
\newc{\eq}[1]{(\ref{eq:#1})}
\newc{\eqs}[2]{(\ref{eq:#1},\ref{eq:#2})}
\newc{\etal}{{\it et al.}\ }
\newc{\ibid}{{\it ibid}.}
\newc{\ibidem}{{\it ibidem}.}
\newc{\eg}{{\it e.g.}\ }
\newc{\ie}{{\it i.e.}\ }
\def \viz{\emph{viz.}}
\def \etc{\emph{etc. }}
\newc{\nonum}{\nonumber}
\newc{\lab}[1]{\label{eq:#1}}
\newc{\dpr}[2]{({#1}\cdot{#2})}
\newc{\lt}{\stackrel{<}}
\newc{\gt}{\stackrel{>}}
\newc{\lsimeq}{\stackrel{<}{\sim}}
\newc{\gsimeq}{\stackrel{>}{\sim}}
\def\lsim{\buildrel{\scriptscriptstyle <}\over{\scriptscriptstyle\sim}}
\def\gsim{\buildrel{\scriptscriptstyle >}\over{\scriptscriptstyle\sim}}
\def\lapp{\mathrel{\rlap{\raise.5ex\hbox{$<$}}
                    {\lower.5ex\hbox{$\sim$}}}}
\def\gapp{\mathrel{\rlap{\raise.5ex\hbox{$>$}}
                    {\lower.5ex\hbox{$\sim$}}}}
\newc{\half}{\frac{1}{2}}
\newcommand {\nnc}        {{\overline{\mathrm N}_{95}}}
\newcommand {\dm}         {\Delta m}
\newcommand {\dM}         {\Delta M}
\def\bra{\langle}
\def\ket{\rangle}
\def\cO#1{{\cal{O}}\left(#1\right)}
\def \DM{{\Delta{m}}}
\newc{\bQ}{\ol{Q}}
\newc{\dota}{\dot{\alpha }}
\newc{\dotb}{\dot{\beta }}
\newc{\dotd}{\dot{\delta }}
\newc{\nindnt}{\noindent}

\newcommand{\medf}[2] {{\footnotesize{\frac{#1}{#2}} }}
\newcommand{\smaf}[2] {{\textstyle \frac{#1}{#2} }}
\def\onesq            {{\textstyle \frac{1}{\sqrt{2}} }}
\def\onehf            {{\textstyle \frac{1}{2} }}
\def\oneth            {{\textstyle \frac{1}{3} }}
\def\twoth            {{\textstyle \frac{2}{3} }}
\def\onefo            {{\textstyle \frac{1}{4} }}
\def\forth            {{\textstyle \frac{4}{3} }}

\newc{\matth}{\mathsurround=0pt}
\def\ML{\ifmmode{{\mathaccent"7E M}_L}
             \else{${\mathaccent"7E M}_L$}\fi}
\def\MR{\ifmmode{{\mathaccent"7E M}_R}
             \else{${\mathaccent"7E M}_R$}\fi}
\newcommand{\s}{\\ \vspace*{-3mm} }

\def \ud { {1 \over 2} }
\def \ut { {1 \over 3} }
\def \td { {3 \over 2} }
\newc{\mr}{\mathrm}
\def\dh {\partial }
\def \cs { cross-section }
\def \css { cross-sections }
\def \cm { centre of mass }
\def \cms { centre of mass energy }
\def \cc { coupling constant }
\def \ccs {coupling constants }
\def \gc {gauge coupling }
\def \gcc {gauge coupling constant }
\def \gccs {gauge coupling constants }
\def \yc {Yukawa coupling }
\def \ycc {Yukawa coupling constant }
\def \pp {{parameter }}
\def \pps {{parameters }} 
\def \ps {parameter space }
\def \pss {parameter spaces }
\def \vv {vice versa }

\newc{\siminf}{\mbox{$_{\sim}$ {\small {\hspace{-1.em}{$<$}}}    }}
\newc{\simsup}{\mbox{$_{\sim}$ {\small {\hspace{-1.em}{$>$}}}    }}


\newc {\Zboson}{{\mathrm Z}^{0}}
\newc{\thetaw}{\theta_W}
\newc{\mbot}{{m_b}}
\newc{\mtop}{{m_t}}
\newc{\sm}{${\cal {SM}}$}
\newc{\as}{\alpha_s}
\newc{\aem}{\alpha_{em}}
\def \PI{{\pi^{\pm}}}
\newc{\ppbar}{\mbox{$p\ol{p}$}}
\newc{\bbbar}{\mbox{$b\ol{b}$}}
\newc{\ccbar}{\mbox{$c\ol{c}$}}
\newc{\ttbar}{\mbox{$t\ol{t}$}}
\newc{\eebar}{\mbox{$e\ol{e}$}}
\newc{\zzero}{\mbox{$Z^0$}}
\def \gamz{\Gam_Z}
\newc{\wplus}{\mbox{$W^+$}}
\newc{\wminus}{\mbox{$W^-$}}
\newc{\ellp}{\ell^+}
\newc{\ellm}{\ell^-}
\newc{\elp}{\mbox{$e^+$}}
\newc{\elm}{\mbox{$e^-$}}
\newc{\elpm}{\mbox{$e^{\pm}$}}
\newc{\qbar}     {\mbox{$\ol{q}$}}
\def \ewgroup{SU(2)_L \otimes U(1)_Y}
\def \smgroup{SU(3)_C \otimes SU(2)_L \otimes U(1)_Y}
\def \smcolorem{SU(3)_C \otimes U(1)_{em}}

\def \SSM  {Supersymmetric Standard Model}
\def \poincare{Poincare$\acute{e}$}
\def \superspace{\emph{superspace}}
\def \sfs{\emph{superfields}}
\def \superpot{\emph{superpotential}}
\def \csf{\emph{chiral superfield}}
\def \csfs{\emph{chiral superfields}}
\def \vsf{\emph{vector superfield }}
\def \vsfs{\emph{vector superfields}}
\newc{\Ebar}{{\bar E}}
\newc{\Dbar}{{\bar D}}
\newc{\Ubar}{{\bar U}}
\newc{\susy}{{{SUSY}}}
\newc{\msusy}{{{M_{SUSY}}}}

\def\photino{\ifmmode{\mathaccent"7E \gam}\else{$\mathaccent"7E \gam$}\fi}
\def\taugluino{\ifmmode{\tau_{\mathaccent"7E g}}
             \else{$\tau_{\mathaccent"7E g}$}\fi}
\def\mphotino{\ifmmode{m_{\mathaccent"7E \gam}}
             \else{$m_{\mathaccent"7E \gam}$}\fi}
\newc{\gl}   {\mbox{$\wt{g}$}}
\newc{\mgl}  {\mbox{$m_{\gl}$}}
\def \charginopm{{\wt\chi}^{\pm}}
\def \mcharginopm{m_{\charginopm}}
\def \mchpmmin {\mcharginopm^{min}}
\def \chonep {{\wt\chi_1^+}}
\def \chone {{\wt\chi_1}}
\def \ch2p {{\wt\chi_2^+}}
\def \chonem {{\wt\chi_1^-}}
\def \ch2m {{\wt\chi_2^-}}
\def \chplus {{\wt\chi^+}}
\def \chminus {{\wt\chi^-}}
\def \chonip{{\wt\chi_i}^{+}}
\def \chonim{{\wt\chi_i}^{-}}
\def \chonipm{{\wt\chi_i}^{\pm}}
\def \chonjp{{\wt\chi_j}^{+}}
\def \chonjm{{\wt\chi_j}^{-}}
\def \chonjpm{{\wt\chi_j}^{\pm}}
\def \chonepm{{\wt\chi_1}^{\pm}}
\def \chonemp{{\wt\chi_1}^{\mp}}
\def \mchonepm{m_{\chonepm}}
\def \mchonemp{m_{\chonemp}}
\def \chtwopm{{\wt\chi_2}^{\pm}}
\def \mchtwopm{m_{\chtwopm}}
\newc{\dmchi}{\Delta m_{\wt\chi}}


\def \vlsp{\emph{VLSP}}
\def \lspi{\wt\chi_i^0}
\def \mlspi{m_{\lspi}}
\def \lspj{\wt\chi_j^0}
\def \mlspj{m_{\lspj}}
\def \lspone{\wt\chi_1^0}
\def \mlspone{m_{\lspone}}
\def \lsptwo{\wt\chi_2^0}
\def \mlsptwo{m_{\lsptwo}}
\def \lspthree{\wt\chi_3^0}
\def \mlspthree{m_{\lspthree}}
\def \lspfour{\wt\chi_4^0}
\def \mlspfour{m_{\lspfour}}


\newc{\sele}{\wt{\mathrm e}}
\newc{\sell}{\wt{\ell}}
\def \msell{m_{\sell}}
\def \slepone{\wt\ell_1}
\def \mslepone{m_{\slepone}}
\def \smuone{\wt\mu_1}
\def \msmuone{m_{\smuone}}
\def \stauone{\wt\tau_1}
\def \mstauone{m_{\stauone}}
\def \snu{\wt{\nu}}
\def \snutau{\wt{\nu}_{\tau}}
\def \msnu{m_{\snu}}
\def \msnumu{m_{\snu_{\mu}}}
\def \barsnu{\wt{\bar{\nu}}}
\def \barsnul{\barsnu_{\ell}}
\def \snul{\snu_{\ell}}
\def \mbarsnu{m_{\barsnu}}
\newc{\snue}     {\mbox{$ \wt{\nu_e}$}}
\newc{\smu}{\wt{\mu}}
\newc{\stau}{\wt{\tau}}
\newc {\nuL} {\wt{\nu}_L}
\newc {\nuR} {\wt{\nu}_R}
\newc {\snub} {\bar{\wt{\nu}}}
\newc {\eL} {\wt{e}_L}
\newc {\eR} {\wt{e}_R}
\def \slepl{\wt{l}_L}
\def \mslepl{m_{\slepl}}
\def \slepr{\wt{l}_R}
\def \mslepr{m_{\slepr}}
\def \stau{\wt\tau}
\def \mstau{m_{\stau}}
\def \slepton{\wt\ell}
\def \mslepton{m_{\slepton}}
\def \mlhiggs{m_{h^0}}

\def \xr{X_{r}}

\def \sfer{\wt{f}}
\def \msfer{m_{\sfer}}
\def \sq{\wt{q}}
\def \msq{m_{\sq}}
\def \msquleft{m_{\tilde{u_L}}}
\def \msqurht{m_{\tilde{u_R}}}
\def \sql{\wt{q}_L}
\def \msql{m_{\sql}}
\def \sqr{\wt{q}_R}
\def \msqr{m_{\sqr}}
\newc{\msqot}  {\mbox{$m_(\sq_{1,2} )$}}
\newc{\sqbar}    {\mbox{$\bar{\wt{q}}$}}
\newc{\ssb}      {\mbox{$\squark\ol{\squark}$}}
\newc {\qL} {\wt{q}_L}
\newc {\qR} {\wt{q}_R}
\newc {\uL} {\wt{u}_L}
\newc {\uR} {\wt{u}_R}
\def \ul{\wt{u}_L}
\def \mul{m_{\ul}}
\newc {\dL} {\wt{d}_L}
\newc {\dR} {\wt{d}_R}
\newc {\cL} {\wt{c}_L}
\newc {\cR} {\wt{c}_R}
\newc {\sL} {\wt{s}_L}
\newc {\sR} {\wt{s}_R}
\newc {\tL} {\wt{t}_L}
\newc {\tR} {\wt{t}_R}
\newc {\stb} {\ol{\wt{t}}_1}
\newc {\sbot} {\wt{b}_1}
\newc {\msbot} {m_{\sbot}}
\newc {\sbotb} {\ol{\wt{b}}_1}
\newc {\bL} {\wt{b}_L}
\newc {\bR} {\wt{b}_R}
\def \mul{m_{\wt{u}_L}}
\def \mur{m_{\wt{u}_R}}
\def \mdl{m_{\wt{d}_L}}
\def \mdr{m_{\wt{d}_R}}
\def \mcl{m_{\wt{c}_L}}
\def \charml{\wt{c}_L}
\def \mcr{m_{\wt{c}_R}}
\newc{\csquark}  {\mbox{$\wt{c}$}}
\newc{\csquarkl} {\mbox{$\wt{c}_L$}}
\newc{\mcsl}     {\mbox{$m(\csquarkl)$}}
\def \msl{m_{\wt{s}_L}}
\def \msr{m_{\wt{s}_R}}
\def \mbl{m_{\wt{b}_L}}
\def \mbr{m_{\wt{b}_R}}
\def \mtl{m_{\wt{t}_L}}
\def \mtr{m_{\wt{t}_R}}
\def \st{\wt{t}}
\def \mst{m_{\st}}
\newc {\stopl}         {\wt{\mathrm{t}}_{\mathrm L}}
\newc {\stopr}         {\wt{\mathrm{t}}_{\mathrm R}}
\newc {\stoppair}      {\wt{\mathrm{t}}_{1}
\bar{\wt{\mathrm{t}}}_{1}}
\def \lstop{\wt{t}_{1}}
\def \lstopbar{\lstop^*}
\def \hstop{\wt{t}_{2}}
\def \hstopbar{\hstop^*}
\def \mlstop{m_{\lstop}}
\def \mhstop{m_{\hstop}}
\def \lstoppair{\lstop\lstop^*}
\def \hstoppair{\hstop\hstop^*}
\newc{\tsquark}  {\mbox{$\wt{t}$}}
\newc{\ttb}      {\mbox{$\tsquark\ol{\tsquark}$}}
\newc{\ttbone}   {\mbox{$\tsquark_1\ol{\tsquark}_1$}}
\def \tsq {top squark }
\def \tsqs {top squarks }
\def \tsql {ligtest top squark }
\def \tsqh {heaviest top squark }
\newc{\mix}{\theta_{\wt t}}
\newc{\cost}{\cos{\theta_{\wt t}}}
\newc{\sint}{\sin{\theta_{\wt t}}}
\newc{\costloop}{\cos{\theta_{\wt t_{loop}}}}
\def \lsbot{\wt{b}_{1}}
\def \lsbotbar{\lsbot^*}
\def \hsbot{\wt{b}_{2}}
\def \hsbotbar{\hsbot^*}
\def \mlsbot{m_{\lsbot}}
\def \mhsbot{m_{\hsbot}}
\def \lsbotpair{\lsbot\lsbot^*}
\def \hsbotpair{\hsbot\hsbot^*}
\newc{\mixsbot}{\theta_{\wt b}}

\def \mhone{m_{h_1}}
\def \hup{{H_u}}
\def \hdn{{H_d}}
\newc{\tb}{\tan\beta}
\newc{\cb}{\cot\beta}
\newc{\vev}[1]{{\left\langle #1\right\rangle}}

\def \abot{A_{b}}
\def \atop{A_{t}}
\def \atau{A_{\tau}}
\newc{\mhalf}{m_{1/2}}
\newc{\mzero} {\mbox{$m_0$}}
\newc{\azero} {\mbox{$A_0$}}

\newc{\lb}{\lam}
\newc{\lbp}{\lam^{\prime}}
\newc{\lbpp}{\lam^{\prime\prime}}
\newc{\rpv}{{\not \!\! R_p}}
\newc{\rpvm}{{\not  R_p}}
\newc{\rp}{R_{p}}
\newc{\rpmssm}{{RPC MSSM}}
\newc{\rpvmssm}{{RPV MSSM}}


\newc{\sbyb}{S/$\sqrt B$}
\newc{\pelp}{\mbox{$e^+$}}
\newc{\pelm}{\mbox{$e^-$}}
\newc{\pelpm}{\mbox{$e^{\pm}$}}
\newc{\epem}{\mbox{$e^+e^-$}}
\newc{\lplm}{\mbox{$\ell^+\ell^-$}}
\def \branch{\emph{BR}}
\def \branche{\branch(\lstop\ra be^{+}\nu_e \lspone)\ti \branch(\lstop^{*}\ra \bar{b}q\bar{q^{\prime}}\lspone)}
\def \branchmu{\branch(\lstop\ra b\mu^{+}\nu_{\mu} \lspone)\ti \branch(\lstop^{*}\ra \bar{b}q\bar{q^{\prime}}\lspone)}
\def\Ecm{\ifmmode{E_{\mathrm{cm}}}\else{$E_{\mathrm{cm}}$}\fi}
\newc{\rts}{\sqrt{s}}
\newc{\rtshat}{\sqrt{\hat s}}
\newc{\gev}{\,GeV}
\newc{\mev}{~{\rm MeV}}
\newc{\tev}  {\mbox{$\;{\rm TeV}$}}
\newc{\gevc} {\mbox{$\;{\rm GeV}/c$}}
\newc{\gevcc}{\mbox{$\;{\rm GeV}/c^2$}}
\newc{\intlum}{\mbox{${ \int {\cal L} \; dt}$}}
\newc{\call}{{\cal L}}
\def \met  {\mbox{${E\!\!\!\!/_T}$}}
\def \cpv  {\mbox{${CP\!\!\!\!/}$}}
\newc{\ptmiss}{/ \hskip-7pt p_T}
\def \eslash{\not \! E}
\def \etslash{\not \! E_T }
\def \ptslash{\not \! p_T }
\newc{\PT}{\mbox{$p_T$}}
\newc{\ET}{\mbox{$E_T$}}
\newc{\dedx}{\mbox{${\rm d}E/{\rm d}x$}}
\newc{\ifb}{\mbox{${\rm fb}^{-1}$}}
\newc{\ipb}{\mbox{${\rm pb}^{-1}$}}
\newc{\pb}{~{\rm pb}}
\newc{\fb}{~{\rm fb}}
\newc{\ycut}{y_{\mathrm{cut}}}
\newc{\chis}{\mbox{$\chi^{2}$}}
\def \hadron{\emph{hadron}}
\def \nlc{\emph{NLC }}
\def \lhc{\emph{LHC }}
\def \cdf{\emph{CDF }}
\def\dzero{\emptyset}
\def \tevatron{\emph{Tevatron }}
\def \lep{\emph{LEP }}
\def \jets{\emph{jets }}
\def \jet(s){\emph{jet(s) }}

\def\Crs{stroke [] 0 setdash exch hpt sub exch vpt add hpt2 vpt2 neg V currentpoint stroke 
hpt2 neg 0 R hpt2 vpt2 V stroke}
\def\loopdk{\lstop \ra c \lspone}
\def\brloopdk{\branch(\loopdk)}
\def\fourdk{\lstop \ra b \lspone  f \bar f'}
\def\brfourdk{\branch(\fourdk)}
\def\fourdklep{\lstop \ra b \lspone  \ell \nu_{\ell}}
\def\fourdkhad{\lstop \ra b \lspone  q \bar q'}
\def\brfourdklep{\branch(\fourdklep)}
\def\brfourdkhad{\branch(\fourdkhad)}
\def\twodk{\lstop \ra b \chonep}
\def\brtwodk{\branch(\twodk)}
\def\threedkslep{\lstop \ra b \wt{\ell} \nu_{\ell}}
\def\brthreedkslep{\branch(\threedkslep)}
\def\threedksnu{\lstop \ra b \wt{\nu_{\ell}} \ell }
\def\brthreedksnu{\branch(\threedksnu) }
\def\threedklsp{\lstop \ra b W \lspone }
\def\brthreedklsp{\\branch(\threedklsp) }
\def\topdk{t \ra \lstop \lspone}
\def\rpvdk{\lstop \ra e^+ d}
\def\brrpvdk{\branch(\rpvdk)}
\def\fonec{f_{11c}} 
\newc{\mpl}{M_{\rm Pl}}
\newc{\mgut}{M_{GUT}}
\newc{\mw}{M_{W}}
\newc{\mweak}{M_{weak}}
\newc{\mz}{M_{Z}}

\newc{\OPALColl}   {OPAL Collaboration}
\newc{\ALEPHColl}  {ALEPH Collaboration}
\newc{\DELPHIColl} {DELPHI Collaboration}
\newc{\XLColl}     {L3 Collaboration}
\newc{\JADEColl}   {JADE Collaboration}
\newc{\CDFColl}    {CDF Collaboration}
\newc{\DXColl}     {D0 Collaboration}
\newc{\HXColl}     {H1 Collaboration}
\newc{\ZEUSColl}   {ZEUS Collaboration}
\newc{\LEPColl}    {LEP Collaboration}
\newc{\ATLASColl}  {ATLAS Collaboration}
\newc{\CMSColl}    {CMS Collaboration}
\newc{\UAColl}    {UA Collaboration}
\newc{\KAMLANDColl}{KamLAND Collaboration}
\newc{\IMBColl}    {IMB Collaboration}
\newc{\KAMIOColl}  {Kamiokande Collaboration}
\newc{\SKAMIOColl} {Super-Kamiokande Collaboration}
\newc{\SUDANTColl} {Soudan-2 Collaboration}
\newc{\MACROColl}  {MACRO Collaboration}
\newc{\GALLEXColl} {GALLEX Collaboration}
\newc{\GNOColl}    {GNO Collaboration}
\newc{\SAGEColl}  {SAGE Collaboration}
\newc{\SNOColl}  {SNO Collaboration}
\newc{\CHOOZColl}  {CHOOZ Collaboration}
\newc{\PDGColl}  {Particle Data Group Collaboration}

\def\issue(#1,#2,#3){{\bf #1}, #2 (#3)}
\def\ASTR(#1,#2,#3){Astropart.\ Phys. \issue(#1,#2,#3)}
\def\AJ(#1,#2,#3){Astrophysical.\ Jour. \issue(#1,#2,#3)}
\def\AJS(#1,#2,#3){Astrophys.\ J.\ Suppl. \issue(#1,#2,#3)}
\def\APP(#1,#2,#3){Acta.\ Phys.\ Pol. \issue(#1,#2,#3)}
\def\JCAP(#1,#2,#3){Journal\ XX. \issue(#1,#2,#3)} 
\def\SC(#1,#2,#3){Science \issue(#1,#2,#3)}
\def\PRD(#1,#2,#3){Phys.\ Rev.\ D \issue(#1,#2,#3)}
\def\PR(#1,#2,#3){Phys.\ Rev.\ \issue(#1,#2,#3)} 
\def\PRC(#1,#2,#3){Phys.\ Rev.\ C \issue(#1,#2,#3)}
\def\NPB(#1,#2,#3){Nucl.\ Phys.\ B \issue(#1,#2,#3)}
\def\NPPS(#1,#2,#3){Nucl.\ Phys.\ Proc. \ Suppl \issue(#1,#2,#3)}
\def\NJP(#1,#2,#3){New.\ J.\ Phys. \issue(#1,#2,#3)}
\def\JP(#1,#2,#3){J.\ Phys.\issue(#1,#2,#3)}
\def\PL(#1,#2,#3){Phys.\ Lett. \issue(#1,#2,#3)}
\def\PLB(#1,#2,#3){Phys.\ Lett.\ B  \issue(#1,#2,#3)}
\def\ZP(#1,#2,#3){Z.\ Phys. \issue(#1,#2,#3)}
\def\ZPC(#1,#2,#3){Z.\ Phys.\ C  \issue(#1,#2,#3)}
\def\PREP(#1,#2,#3){Phys.\ Rep. \issue(#1,#2,#3)}
\def\PRL(#1,#2,#3){Phys.\ Rev.\ Lett. \issue(#1,#2,#3)}
\def\MPL(#1,#2,#3){Mod.\ Phys.\ Lett. \issue(#1,#2,#3)}
\def\RMP(#1,#2,#3){Rev.\ Mod.\ Phys. \issue(#1,#2,#3)}
\def\SJNP(#1,#2,#3){Sov.\ J.\ Nucl.\ Phys. \issue(#1,#2,#3)}
\def\CPC(#1,#2,#3){Comp.\ Phys.\ Comm. \issue(#1,#2,#3)}
\def\IJMPA(#1,#2,#3){Int.\ J.\ Mod. \ Phys.\ A \issue(#1,#2,#3)}
\def\MPLA(#1,#2,#3){Mod.\ Phys.\ Lett.\ A \issue(#1,#2,#3)}
\def\PTP(#1,#2,#3){Prog.\ Theor.\ Phys. \issue(#1,#2,#3)}
\def\RMP(#1,#2,#3){Rev.\ Mod.\ Phys. \issue(#1,#2,#3)}
\def\NIMA(#1,#2,#3){Nucl.\ Instrum.\ Methods \ A \issue(#1,#2,#3)}
\def\JHEP(#1,#2,#3){J.\ High\ Energy\ Phys. \issue(#1,#2,#3)}
\def\EPJC(#1,#2,#3){Eur.\ Phys.\ J.\ C \issue(#1,#2,#3)}
\def\RPP (#1,#2,#3){Rept.\ Prog.\ Phys. \issue(#1,#2,#3)}
\def\PPNP(#1,#2,#3){ Prog.\ Part.\ Nucl.\ Phys. \issue(#1,#2,#3)}
\newc{\PRDR}[3]{{Phys. Rev. D} {\bf #1}, Rapid  Communications, #2 (#3)}

\vspace*{\fill}
\vspace{-1.5in}
\begin{flushright}
{\tt IISER-K/HEP/06/12}
\end{flushright}
\begin{center}
{\Large \bf New limits on top squark NLSP from \\
 ATLAS 4.7 $fb^{-1}$ data }
  \vglue 0.4cm
  Arghya Choudhury\footnote{arghyac@iiserkol.ac.in} and
  Amitava Datta\footnote{adatta@iiserkol.ac.in}
      \vglue 0.1cm
          {\it 
	  Indian Institute of Science Education and Research - Kolkata, \\
          Mohanpur Campus, PO: BCKV Campus Main Office,\\
          Nadia, West Bengal - 741252, India.\\
	  }
	  \end{center}
	  \vspace{.1cm}

\vspace{+1cm}
\begin{abstract}
Using the ATLAS 4.7 $\ifb$ data on new physics search in the jets + $\met$ channel,
we obtain new limits on the lighter top squark ($\tilde t_1$) considering all its decay modes 
assuming that it is the next to lightest supersymmetric particle (NLSP). 
If the decay $\lstop$ $\ra c \lspone$  dominates and the production of dark 
matter relic density is due to NLSP - LSP co-annihilation 
then the lower limit on  $\mlstop $ is 240 GeV. 
The limit changes to 200 GeV  if the decay $\lstop$ $\ra b W \lspone$ 
dominates. Combining these results it follows that $\lstop $ 
NLSP induced baryogenesis is now constrained more tightly.

\vspace{4 cm}

\end{abstract}

Models with Supersymmetry (SUSY) \cite{revSUSY1,revSUSY2,revSUSY3,revSUSY4} are very well motivated extensions 
of the Standard model (SM) of particle physics. 
The LHC experiments \cite{atlas0l,cms} with 4.7 $\ifb$ data have severely constrained the masses of 
the strongly interacting supersymmetric particles (sparticles) - the 
squarks and the gluinos \cite{revSUSY1,revSUSY2,revSUSY3,revSUSY4}. However, a subset of these 
sparticles may still be rather light. An especially interesting 
possibility is the light stop scenario. In this case only one of the  squarks - 
the lighter superpartner of the top quark or stop ($\lstop$) - is light and within the reach 
of the ongoing LHC experiments along with the electroweak sparticles (the 
sleptons and the electroweak gauginos)\footnote{ It has recently been 
emphasized \cite{arghya1,arghya2,arghya3} that if the squarks and the gluinos 
are compatible with the LHC mass bounds, the data is consistent with a 
wide variety of relatively light electroweak sparticles and they may 
have masses just above the corresponding lower limits from LEP 
\cite{lepsusy}.}.

The light stop scenario is theoretically well-motivated. Due to a 
large mixing term - driven by the top quark mass - in the stop mass 
matrix, one of the mass eigenvalues (i.e $\mlstop$) may turn out to be much smaller 
than the masses all other squarks. This result is model independent. 
Additional suppression of $\mlstop$ may arise due to the underlying 
SUSY breaking mechanism. In the minimal supergravity (mSUGRA) 
model \cite{msugra1,msugra2,msugra3,msugra4,msugra5}, e.g., all squarks have a common mass at the Grand  
Unification Theory (GUT) scale ($M_G$). Yet  $\mlstop$ at the weak scale may be suppressed 
relative to the masses of the other squarks due to renormalization group running between 
$M_G$ and the weak scale driven by the large top quark Yukawa coupling. 
In this paper,however, we consider the 
light stop scenario at the weak scale as a phenomenological model without 
invoking any specific mechanism for the suppression of $\mlstop$. 
Light stop squarks are cosmologically interesting 
since they can provide an explanation of baryogenesis \cite{baryo1,baryo2,baryo3,baryo4,baryo5}. 
This happens if 120 $\lsim \mlstop \le m_t $. 

If $\lstop$ is the next to lightest super particle (NLSP) then another 
cosmologically interesting possibility opens up. The observed dark matter
(DM) \cite{dmreva,dmrevb,dmrevc,dmrev1} relic density of the universe \cite{wmap} may be produced
by the co-annihilation of $\lstop$ with the lightest supersymmetric particle (LSP),   
the DM candidate \cite{lspstopcoan1,lspstopcoan2}. We shall consider models where 
the lightest neutralino ($\lspone$) is the LSP. 
For this co-annihilation, however, $\Delta m = \mlstop - \mlspone$ has to be rather small.
The novel idea that the light stop scenario
plays the  major role in both baryogenesis and DM relic density production
has also been emphasized in the literature \cite{carena1,carena2}.   
In this work we shall focus on the possible constraints on the stop NLSP scenario 
for both large and small $\Delta m$ from the published LHC 7-TeV data 
in the jets + $\met$ channel \cite{atlas0l} 
corresponding to integrated luminosity $\lum$ = 4.7 $\ifb$. 
Similar analyses were done for $\lum$ = 1 $\ifb$ data in \cite{stopnlsp1,stopnlsp2,stopnlsp3}

Before starting the main discussion it is worthwhile to review the existing bounds on the stop
NLSP from the LEP and the Tevatron. The Stop NLSP has the following decay modes:\\ 
(i) $\lstop$ $\ra c \lspone$ (the loop induced mode) \cite{hikasa1,hikasa2,djouadi}, 
(ii) $\lstop$ $\ra b W \lspone$, (iii) $\lstop$ $\ra t \lspone$ and (iv) $\lstop$ $\ra b f \bar {f^{\prime}} \lspone$

The Branching ratios of the modes (i) and (iv) are negligible if either (ii) or (iii) is kinematically accessible.
The modes (i) and (iv) can in principle compete with one another \cite{djouadi}, but the latter
is negligible over most of the parameter space if tan $\beta$ is moderately 
large ($\geq$ 7) \cite{djouadi,admgsp} and will be ignored in this paper.
If the $\lstop - \lspone$ co-annihilation is indeed responsible for the DM relic density 
production then only decay mode (i) is allowed. 
In fact the signal arising due to mode (ii) has not received the due attention in the 
literature.

Most of the existing bounds on stop NLSP are derived by assuming that the loop 
induced decay (i) occur with 100\%. The expected signal is (at least) 2 jets + $\met$. 
For $\mlstop$ within  the kinematical reach of LEP, the bound is 
sensitive to small values of $\Delta m$ as well and is given by $\mlstop$ $\ge$ 100 (95.2) GeV for 
$\Delta m \ge$ 20 (5) GeV for no mixing ($\theta_{\tilde t}$ = 0) \cite{lepsusy}. 
The numerical change in the limit is not drastic for other values of $\theta_{\tilde t}$. 
On the other hand the Tevatron experiments are sensitive to larger $\mlstop$ but 
are valid for rather larger $\Delta m$. 
The best bound  is $\mlstop$ $\ge$ 180 GeV for $\mlspone = 90$ GeV \cite{cdf}. 
Thus stop induced baryogenesis is disfavoured for $\Delta m$ $\simeq$ 90 GeV. 
For smaller $\Delta m$ the efficiency of the kinematical selection requiring high $p_T$ jets 
becomes rather poor due to the modest energy release in the decay
process and the limit becomes weaker. 
The possibility of stop NLSP induced baryogenesis, therefore, remains open for small $\Delta m$. 
It is, also, fair to conclude that the possibility of $\lstop$ - LSP co-annihilation is only
excluded for $\mlstop$ $\lsim$ 100 from the LEP bound.

The ATLAS or the CMS collaboration has not yet published any dedicated search for the pair production
of the stop NLSP\footnote{After the completion of our paper ATLAS collaboration has published 
limits on $\mlstop$  based on the  $\lstop$ $\ra t \lspone$ mode using different cuts \cite{atlasstop1,atlasstop2,atlasstop3}.}. 
However, this pair production followed by the decay of each stop in channels i) - iv)
would certainly contribute to the jets + missing $E_T$ signature carefully looked into by the LHC collaborations. 
The non-observation of this signal can potentially constrain the light stop scenario. 
It has, already been noted that the $\lum \approx 1 \ifb$ data was sensitive 
to the stop sector \cite{stopnlsp1,stopnlsp2,stopnlsp3}. 
However, we shall show below that the limits are much stronger for the $\lum$ = 4.7 $\ifb$ data. 
This large $\lum$ accumulated at the LHC enables one to constrain the stop NLSP scenario meaningfully 
for smaller $\Delta m$  and directly address the issue of Stop NLSP - LSP co-annihilation as a DM 
producing mechanism at a hadron collider.

We have computed the sparticle spectra and decay branching ratios (BRs) 
in the minimal supersymmetric extension of the standard model (MSSM) 
without any assumption regarding the soft breaking parameters 
using SUSPECT (v2.41) \cite{suspect} and SDECAY \cite{sdecay}. 
Using PYTHIA (v6.424)\cite{pythia} Monte Carlo (MC) event generator we have 
generated the jets + $\met$  signal stemming from  $\lstop \lstop^*$ events only. 
The next to leading order (NLO) cross-section for the $\lstop \lstop^*$ pair
production have been computed by PROSPINO 2.1 \cite{prospino} with CTEQ6.6M  PDF \cite{cteq6.6}.   
The normalization and factorization scale is chosen to be $\mlstop$. 
We have taken the pole mass of the top quark (running bottom 
quark mass evaluated in the $\overline{MS}$ scheme) as $m_t$ $(m_b)=$ 173.2 (4.25).
We have computed the DM relic density using micrOMEGAs (v.2.4.1) \cite{micromegas}. 
The observed DM relic density ($\Omega h^2$) in the universe measured by the WMAP collaboration \cite{wmap}
is given by $\Omega h^2$ = 0.1126 $\pm$ 0.0036. 
If 10$\%$ theoretical uncertainty is added \cite{baro} then the 
DM relic density is bounded by 0.09 $\leq \Omega h^2\leq$ 0.13 at 2$\sigma$ level.

\begin{figure}[!htb]
\begin{center}
\includegraphics[angle =0, width=1.0\textwidth]{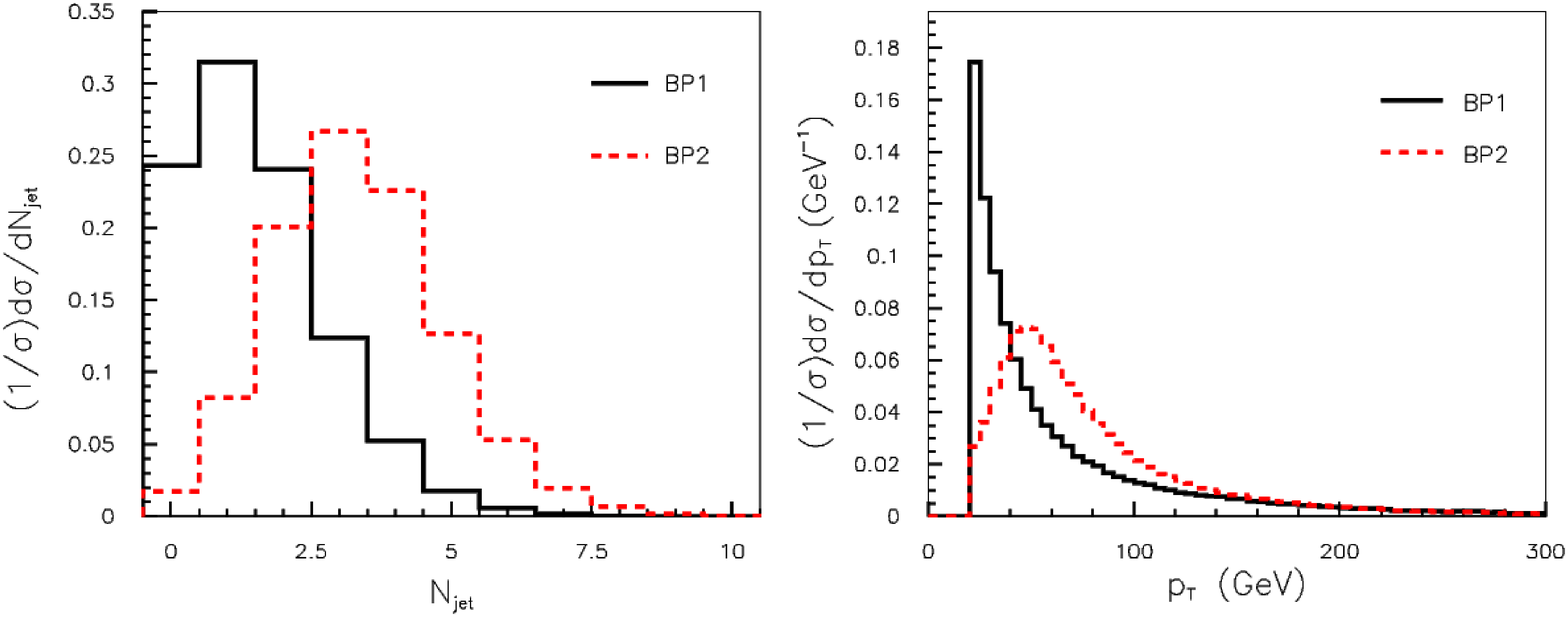}
\end{center}
\caption{ 
{\footnotesize The distribution of $N_{jet}$ (in the left panel) and $p_{T}$ (in the right panel) of the hardest 
jet for the Benchmark points BP1 and BP2 (see text for details).   } }
\end{figure}

ATLAS Collaboration has updated their results for SUSY searches in $jets + \met + 0l $ 
final state for $\lum$ = 4.7 $\ifb$ \cite{atlas0l}. 
They have defined five inclusive analysis channels labelled as A to E according to jet multiplicity ($N_j$ = 2 to 6). 
Further, the  $N_j$ = 2 sample is divided into two channels - A and A'. The channels 
and the details of kinemetical cuts for each of them are given in Table 1 of \cite{atlas0l}.
Depending upon the final cuts on the observable $m_{eff}$(incl.) (defined as the scalar sum of all jets with $P_T  >$  40 GeV 
and $\met$) each channel is further classified as `Tight',`Medium' and `Loose'.  In this way they 
have finally presented the results for 11 signal regions and have constrained 
any new physics model in terms of upper limit on the number of events $N_{BSM}$ or effective cross-section $\sigma_{BSM}$/fb. 
The observed upper limits on $N_{BSM}$ at  
95 $\%$ Confidence Level (CL) for signal regions SRA-Tight, SRA-Medium, SRA'-Medium, SRB-Tight, SRC-Tight, 
SRC-Medium, SRC-Loose, SRD-Tight, SRE-Tight, SRE-Medium, SRE-Loose are 
2.9, 25, 29, 3.1, 16, 18, 58, 10, 12, 12, 84 respectively \cite{atlas0l}.


We have adopted the selection criteria for different 
signal regions used by the ATLAS Collaboration \cite {atlas0l}. Then we have checked whether 
the number of events from the $\lstop \lstop^*$ pair production followed by the decays (i) - (iii)  
of both the squarks exceed the corresponding upper bound for at least one of  
the signal regions to obtain the 95 $\%$ CL exclusion contour in the  $\mlstop$ - $\mlspone$ plane.

If the BR of the decay mode $\lstop$ $\ra c \lspone$  is 100 $\%$ then the signal will be 
2j + $\met$. In this case SRA-Medium or SRA'-Medium, specially designed for compressed 
spectra \cite{atlas0l} is very effective in excluding the 
regions of the $\mlstop - \mlspone$ space.  
For small $\Delta m $, SRA'-Medium is the most sensitive one. 
On the otherhand, high jet multiplicity regions chosen by ATLAS 
are optimised for long decay chains \cite{atlas0l}.  For $\lstop$ $\ra b W \lspone$ or $t \lspone$ 
decay, events will contain at least 2, 4 or 6 jets. Due to the lepton veto 2j signal will not be sensitive at 
all. In our analysis we have also checked that  when BR of $\lstop$ $\ra t \lspone$ is 100 $\%$ 
then the exclusion mainly comes from SRE-Medium ($N_j = 6 $) channel. Again for $\lstop$ $\ra b W \lspone$ 
effective signal regions are either SRC-Medium or SRE-Medium. 
The distributions of the number of jets ($N_j$) and the $p_T$ of the hardest jet in the signal 
are shown in Fig. 1 for the two representative point BP1 and BP2. 
For both of them $\mlstop$ is 200 GeV. But $\mlspone$ = 180 and 100 for BP1 and BP2 
respectively. 
Both the distributions are drawn with nominal cuts and normalized to unity. 

\begin{figure}[!htb]
\begin{center}
\includegraphics[angle =270, width=1.0\textwidth]{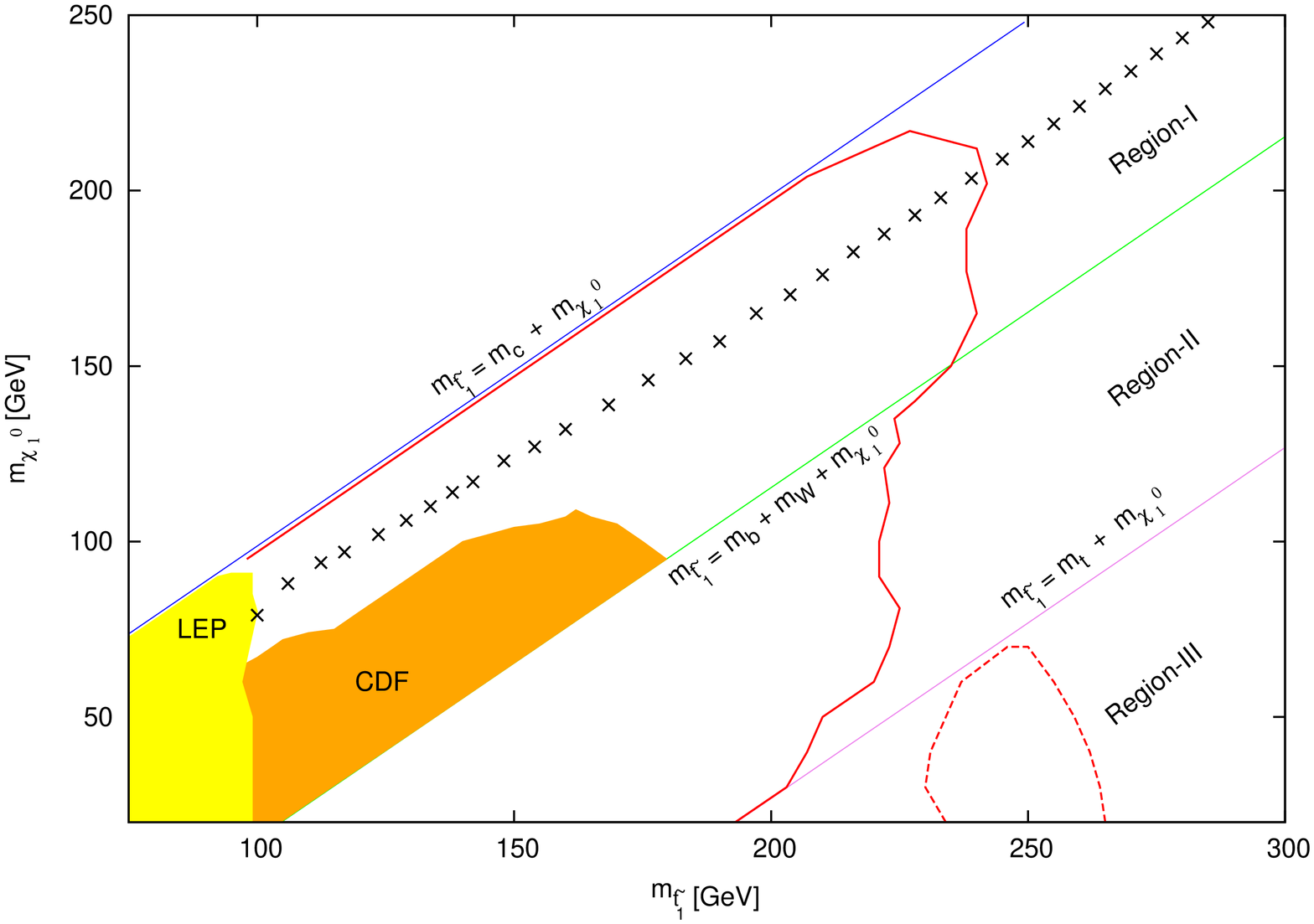}
\end{center}
\caption{ {\footnotesize Exclusion limits  at 95 $\%$ CL in the  $\mlstop$ - $\mlspone$ plane 
for different decay modes of the $\lstop$ - NLSP. 
The x marked points represent the parameter space where $\lstop - \lspone$ co-annihilation 
is viable.   }}
\end{figure}

Fig. 2 depicts the regions in the $\mlstop$ - $\mlspone$ plane excluded by the ATLAS jets + $\met$ 
data. In the region between the $\mlstop = m_c + \mlspone$ and the $\mlstop = m_b + m_W + \mlspone$ 
lines (Region-I), the loop decay (i) dominates. 
We have restricted ourselves to $(\mlstop)_{min} > m_c + \mlspone$ + 2 GeV. 
The same increment of 2 GeV has been introduced at each point 
representing a cross over from one region to another.
In Region-II bounded by the  lines $\mlstop = m_b + m_W + \mlspone$ 
and $\mlstop = m_t + \mlspone$ lines, the three body decay channel (ii) dominates. 
In regions I and II the parameter space to the left of the curved line is excluded. 
In Region-III  below the line  $\mlstop = m_t + \mlspone$ , the two body decay 
mode (iii) is the only significant decay mode. The parameter space bounded by the dashed line and the x-axis 
is excluded.
The following points may be noted :

a) For very small $\Delta m$ the jets in the signal essentially come from initial state and final state 
radiation. This result is, therefore , crucially dependent  on the parton showering model 
in Pythia. In contrast for $\Delta m$ = 25 - 30 GeV, which is the case in 
the co-annihilation region, even the parton level $p_T$ distribution 
develops a long tail. 
Our results for this region is seem to be more reliable. 
In the co-annihilation region $\mlstop$ $\lsim$ 240 GeV is disfavoured. 
In Region-II the best limit is  $\mlstop$ $\gsim$  230 GeV for $\mlspone$ $\simeq$ 150 GeV. 
In region III there is no unambiguous lower limit on $\mlstop$ but a patch in the parameter 
space is disfavoured. It may, however, be recalled that no significant constraint on $\mlstop$ 
exists for region II and III. 

b) The $\lstop$ NLSP induced baryogenesis requiring 120 $\lsim \mlstop \le m_t $ seems to be disfavoured 
in the co-annihilation region. Consequently  the scenario where both DM relic density 
and baryogenesis are induced by the $\lstop$ NLSP is rather unlikely. 
In Region-II $\mlstop$ $\gsim$ 200 GeV and hence, $\lstop$ induced 
baryogenesis is also disfavoured. In Region-III it is forbidden by definition. 

c) Significant constraints are obtained in decay channels (ii) and (iii).
The strong constraints obtained even for small $\Delta m$ are due to the large $\lum$ 
accumulated at the current LHC 7 TeV experiments. 
Most of the above bounds come from ATLAS selection criteria requiring two high $p_T$ 
jets in the final state . That is why the constraints in Region-III are rather modest. 
Here the final state contains either large number of jets and/or at least one high $p_T $ lepton which is vetoed.


\newpage

\end{document}